\documentclass[superscriptaddress,twocolumn,showpacs,pra,longbibliography]{revtex4-2}
\usepackage{bm} 
\usepackage{graphicx} 
\usepackage{amsmath,amsthm,amssymb,amsfonts}
\usepackage{comment}
\usepackage{color} 
\usepackage{mathtools}
\usepackage{MnSymbol}
\usepackage{hyperref}
\usepackage{multirow}
\usepackage{qcircuit}
\hypersetup{
    colorlinks=true,       
    linkcolor=cyan,          
    citecolor=magenta,        
    filecolor=magenta,      
    urlcolor=cyan,           
    runcolor=cyan
}

\newcommand{\ket}[1]{| #1 \rangle}

\begin{document}

\title{A Post-Training Approach for  Mitigating Overfitting in Quantum Convolutional Neural Networks}

\author{Aakash Ravindra Shinde}
\affiliation{Graduate School,  Viterbi School of Engineering, University of Southern California, University of Southern California, Los Angeles, California 90089, USA}
\author{Charu Jain}
\affiliation{Graduate School,  Viterbi School of Engineering, University of Southern California, University of Southern California, Los Angeles, California 90089, USA}
\author{Amir Kalev}
\affiliation{Information Sciences Institute, University of Southern California, Arlington, VA 22203, USA}
\affiliation{Department of Physics and Astronomy, and Center for Quantum Information Science \& Technology, University of Southern California, Los Angeles, California 90089, USA}
\begin{abstract}
Quantum convolutional neural network (QCNN), an early application for quantum computers in the NISQ era, has been consistently proven successful as a machine learning (ML) algorithm for several tasks with significant accuracy. Derived from its classical counterpart, QCNN is prone to overfitting. Overfitting is a typical shortcoming of ML models that are trained too closely to the availed training dataset and perform relatively poorly on unseen datasets for a similar problem.  In this work we study post-training approaches for mitigating overfitting in QCNNs. We find that a straightforward adaptation of a classical post-training method, known as neuron dropout, to the quantum setting leads to a significant and undesirable consequence: a substantial decrease in success probability of the QCNN. We argue that this effect exposes the crucial role of entanglement in QCNNs and the vulnerability of QCNNs to entanglement loss. Hence, we propose a parameter adaptation method as an alternative method. Our method is computationally efficient and is found to successfully handle overfitting in the test cases.
\end{abstract}
\maketitle

\section{Introduction}
Quantum machine learning (QML) has shown great promise on several prototypes of NISQ devices, presenting significant accuracy over several different datasets, see e.g., ~\cite{abohashima2020classification,schatzki2021entangled} and references therein. It has been proven effective even on a limited number of available qubits, not only on simulated devices but even when tested on noise-prone quantum information processing devices, as seemingly proven difficult in the case presented in~\cite{mancilla2022preprocessing}. As most of QML models have been derived from the classical machine learning (ML) models, with adaptations over the functioning being operable on quantum devices, these algorithms pose similar challenges as found in their classical counterparts. 
One of the setbacks of ML models, classical and quantum included, is overfitting, causing models to underperform on average when presented with external data  outside the training set for the same problem. Due to its importance, the problem of overfitting has been studied extensively in the classical ML literature, and several methods have been proposed and implemented to mitigate it, including data augmentation, regularization, early stopping, and neuron dropout~\cite{srivastava2014dropout}. A recent review of the state-of-the-art techniques to handle overfitting in the classical setting can be found, for example in Ref.~\cite{ying2019overview}.


In contrast to the classical setting, there has been much less investigation as to how to address the problem of overfitting in QML models.  Nevertheless, the problem of overfitting has been recently recognized and investigated in the context of QML, with significant results presented in~\cite{verdon2018universal, schuldCircuit2020, chenExpress2021,Kobayashi2022,peters2023generalization}. In Ref.~\cite{peters2023generalization}
the authors discuss characterization of benign overfitting in quantum models through the lens of the structure of quantum circuits, such as  data-encoding and state preparation. In addition, inspired by the  dropout technique that implemented to handle overfitting in the classical setting, the authors of~\cite{Kobayashi2022} propose a CNOT gate dropout technique and show  its effectiveness on a few trial quantum regression models and test cases. Earlier works~\cite{verdon2018universal, schuldCircuit2020} also considered qubit dropout as an overfitting mitigation technique in the context of quantum learning, and show its applicability with promising results.



Most, if not all, techniques that have been proposed for suppressing overfitting in QML models are performed  during training. In this work we propose a post-training method, complementing existing works. The motivation and advantage of considering post-training methods to handle overfitting in QML models is twofold. Firstly, whereas during-training techniques require modification of the quantum circuit structure from one training step to another (for example, by dropping out gates and qubits), in post-training methods the circuit structure is fixed during the costly phase of training. In addition to this, post-training methods can be used in conjunction of during-training methods to ensure that the performance of the final model is numerically optimized.  In general terms, once a model has been trained, our technique takes into account the values that each trained parameter takes on in the last few iterations (a number which can be treated as a hyper-parameter and can be numerically optimized), and based on these values assigns a new value to each parameter (e.g., their mean).  As such, the proposed technique is computationally efficient, requires no additional training, and can be readily combined with pre- and during-training methods to enhance the performance of QML models on unseen data. We refer to our method as PTA (Post-Training parameter Adjustment). For concreteness, and since it is one of the widely-used QML architectures, we showcase our techniqe and test it using quantum convolutional neural networks (QCNNs)~\cite{cong2019quantum}.

The paper is organized as follows: In Sec.~\ref{sec:soft} we present the various methods and techniques we used and developed in this work.   In Sec.~\ref{sec:results} we present and discuss our numerical experimental results. We offer conclusions and outlook in Sec.~\ref{sec: Conclusion}.

\section{Methods and Techniques}\label{sec:soft}
Before presenting the results from our numerical experiments, we devote this section to providing an overview of the main tools and techniques we used and developed in this work. 

\subsection{The QCNN architecture}\label{subsec:arch}
QCNNs are essentially variational quantum algorithms~\cite{cong2019quantum,hur2022quantum}. Similar to their classical counterparts, QCNN is designed and used for solving classification problems (supervised and unsupervised paradigms have been studied in this context)~\cite{perelshtein2022practical}.  They were proposed to be well-fitted for NISQ computing due to their intrinsically shallow circuit depth.  It was shown that due to unique quantum phenomena, such as superposition and entanglement, QCNN can provide better prediction statistics using less training data than classical ones in certain circumstances~\cite{reese2022predict}.

Due to the noise and technical challenges of building quantum hardware, the size of quantum circuits that can be reliably executed on NISQ devices is limited. Thus, the encoding schemes for high dimensional data usually require a number of qubits that are beyond the current capabilities of quantum devices. Therefore, classical dimensionality reduction techniques are particularly useful in the near-term application of QML techniques. In this work, the classical data was pre-processed using two dimensionality reduction techniques, namely Principal Component Analysis (PCA)~\cite{jolliffe2002principal} and Autoencoding (AutoEnc)~\cite{goodfellow2016deep}. Autoencoders are capable of modeling complex non-linear functions, whereas PCA is a simpler linear transformation that helps in cheaper and faster computation.

\begin{figure*}[!ht]
  \centering
  \includegraphics[width=0.95\textwidth]{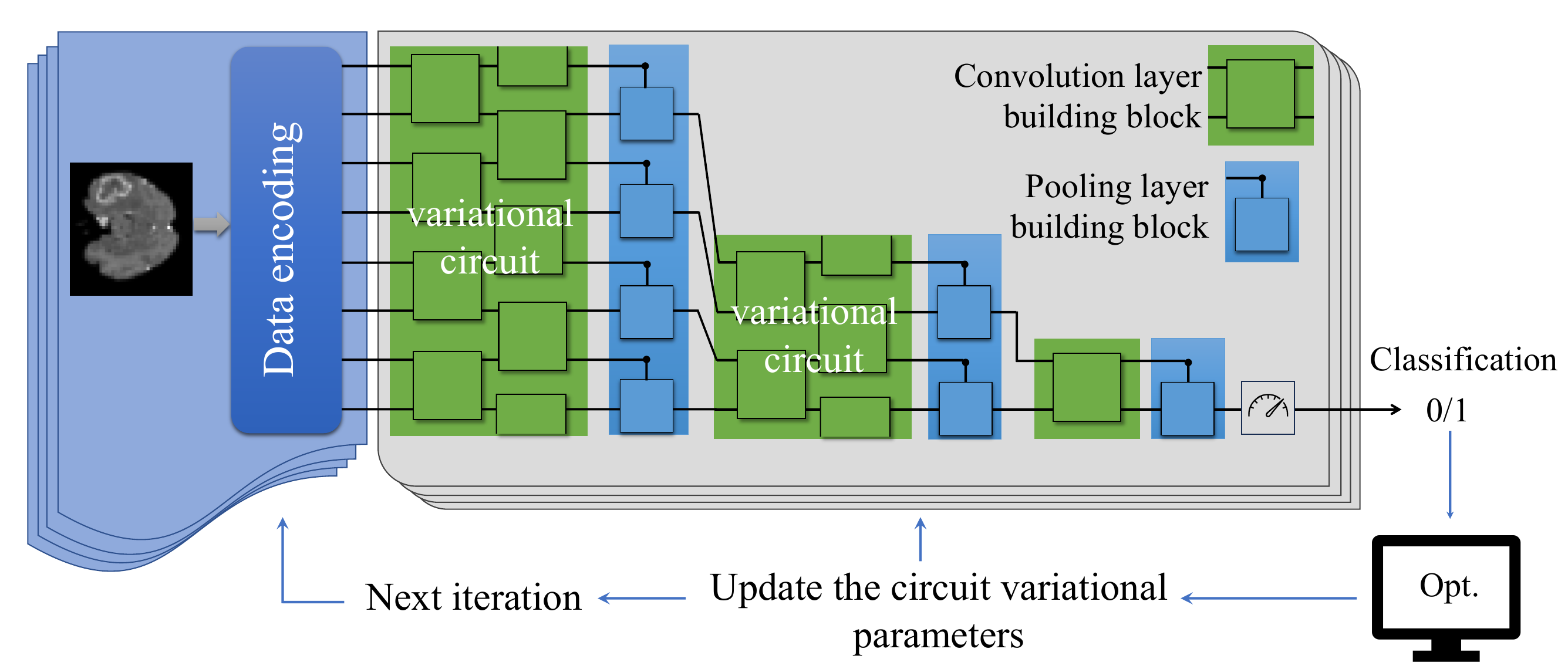} 
  \caption{{\bf General QCNN architecture}. The QCNN includes three key components: A feature map, a parametric quantum circuit that includes concatenated convolution and pooling layers, and a measurement followed by an optimization unit. In this work the convolution and the pooling layers are constructed from two-qubit gates (building blocks). Examples of the building blocks we used are given in Fig.~\ref{fig:conv}.}
  \label{fig:qml_setup}
\end{figure*}
A generic QCNN is composed of a few key components~\cite{cong2019quantum,hur2022quantum}, as illustrated in Fig.~\ref{fig:qml_setup}. The first component is data encoding, also known as a quantum feature map. In classical ML, feature maps are used to transform input data into higher-dimensional spaces, where the data can be more easily separated or classified. Similarly, a quantum feature map transforms classical data into a quantum state representation. The main idea is to encode the classical data as an entangled state with the possibility of capturing richer and more complex patterns within the data. The quantum feature map is done in practice by applying a unitary transformation to the initial state (typically the all-zero state). In this work, we implemented two of the main feature encoding schemes, amplitude encoding and qubit encoding~\cite{cong2019quantum,hur2022quantum}. In the former,  classical data $(x_1,\ldots,x_{k})\in \mathrm{R}^{k}$ is represented as, generally,  an entangled input quantum state $\ket{\psi_{\rm in}}\sim\sum_{i=1}^k x_i \ket{i}$ (up to normalization), where $\ket{i}$ is a computational basis ket. Amplitude encoding uses a circuit depth of size ${\cal O}(\log N)$ circuit and  $N$ qubits~\cite{araujo2021divide}. To evaluate the robustness of our dropout method with respect to the feature map, we also used qubit encoding. In this method the input state is a separable state $\ket{\psi_{\rm in}}=\bigotimes_{i=1}^k (\cos\frac{x_i}{2}\ket {0} + \sin\frac{x_i}{2}\ket{1})$. As such, it uses a constant-depth circuit given by a product of a single-qubit rotation.

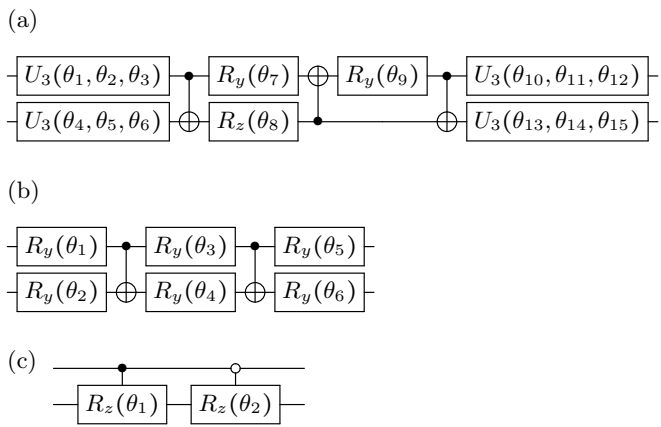
\begin{figure}[!ht]
\flushleft{(a)}
\begin{center}
\Qcircuit @C=0.4em @R=.3em {
& \gate{U_3(\theta_1, \theta_2, \theta_3)} & \ctrl{1} & \gate{R_y(\theta_7)} & \targ & \gate{R_y(\theta_9)} & \ctrl{1} & \gate{U_3(\theta_{10}, \theta_{11}, \theta_{12})} & \qw \\
& \gate{U_3(\theta_4, \theta_5, \theta_6)} & \targ & \gate{R_z(\theta_8)} & \ctrl{-1} & \qw &\targ & \gate{U_3(\theta_{13}, \theta_{14}, \theta_{15})} & \qw
}
\end{center}
\flushleft{(b)}
\begin{center}
\Qcircuit @C=0.4em @R=.3em {
& \gate{R_y(\theta_1)} & \ctrl{1} & \gate{R_y(\theta_3)} & \ctrl{1} & \gate{R_y(\theta_5)}  & \qw \\
& \gate{R_y(\theta_2)} & \targ & \gate{R_y(\theta_4)} & \targ & \gate{R_y(\theta_6)} & \qw
}
\end{center}
\begin{flushleft}{(c)}
~\Qcircuit @C=1.0em @R=0.5em {
& \ctrl{1} & \ctrlo{1} & \qw \\
& \gate{R_z(\theta_1)} & \gate{R_z(\theta_2)} & \qw
}
\end{flushleft}
\caption{{\bf Two-qubit building blocks of the implemented QCNN.} We used the architecture proposed and implemented in~\cite{hur2022quantum}. The building blocks for the convolution layers are given in subfigure (a) and (b) where $U_3(\theta, \varphi, \lambda) = R_z(\varphi)R_x(-\frac{\pi}{2})R_z(\theta)R_x(\frac{\pi}{2})R_z(\lambda)$, while the building block for the pooling layer is shown in subfigure (c).}
\label{fig:conv}
\end{figure}

The second key component of a QCNN is a parameterized quantum circuit (PQC)~\cite{sim2019expressibility,benedetti2019Quantum}. PQCs are composed of quantum gates whose action is determined by the value of a set of parameters. Using a variational algorithm (classical or quantum), the PQC is trained by optimizing the parameters of the gates to yield the highest accuracy in solving the ML task (e.g., classification) on the input data. Typically, in QCNN architectures, the PQC is composed of a repeated sequence of a (parametric) convolution circuit followed by and a (parametric) pooling circuit. The convolution layer is used as the PQC for training a tree tensor network (TTN)~\cite{grant2018hierarchical}. In this work, we used a specific form of the convolution layer, which was proposed and implemented by Hur {\it et al.}~\cite{hur2022quantum}, and that is constructed out of a concatenation of two-qubit  gates (building blocks). In Fig.~\ref{fig:conv}(a)-(b) we sketch two of the building blocks that  we used for convolution layer in our architecture.  

The convolution layer is followed by a pooling layer, which reduces the dimensionality of the input data while preserving important features, i.e., the pooling layer applies parameterized (controlled) quantum gates on the sequence of two qubits. To reduce the dimensionality, the control qubits are traced out (assuming they maintain coherence through the computation) while the target qubits continue to the next convolution layer, see Fig.~\ref{fig:qml_setup}. For the implementation of the pooling layer, we used a parameterized two-qubit circuit that consisting of two controlled rotations $R_z (\theta_1)$ and $R_x (\theta_2)$, respectively, each activated when the control qubit is 1  or 0 (filled and open circle in Fig.~\ref{fig:conv}(c)). The PQC is followed by a measurement in the computational basis on the last layer of qubits.

Training the QCNN is obtained by successively optimizing the PQC using the input data and their labels by minimizing an appropriate cost function. Here we use the mean squared error (MSE) between predictions and class labels. Given a set of training data $\{\ket{\psi_i}, y_i\}$ of size $K$, where $\ket{\psi_i}$ denotes an initial state and $y_i\in \{0,1\}$ denotes their label, the MSE cost function is given by
\begin{align}
C(\boldsymbol{\theta})=\frac1{K}\sum_{i=1}^K\Big(y_i-f_{\boldsymbol{\theta}}(\ket{\psi_i})\Big)^2.
\end{align}
Here, $f_{\boldsymbol{\theta}}(\ket{\psi_i})$ is the output of QCNN ($f\in\{0,1\}$) which depends on the set of parameters $\boldsymbol{\theta}$ that define the gates of the PQC. 

\subsection{Post-training gate dropout}\label{subsec:dropout}
Inspired by the Monte-Carlo dropout technique (a popular classical technique that uses neuron dropout for uncertainty quantification)~\cite{gal16dropout} and the recent  work and results by Kobayashi {\it et al}.~\cite{Kobayashi2022} we have tested a  method in which a small fraction gates -- single-qubit (parametric) gates or CNOT gates -- are dropped out from the circuit after training,  equivalently, replacing a gate with the identity gate.  None of the controlled two-qubit variational gates in the pooling layers were dropped out. As we discussed in length in  Sec.~\ref{sec:results}, we found that this dropout method fails catastrophically. Not only did it not help with mitigating overfitting, it substantially reduced the success rates on the {\it trained} data. The results seems to be persistent irrespective to the feature map, to the structure of the variational building block circuit, and to the data set used in the numerical experiments. We report this ``negative'' result since we believe,  it can provide a better understanding onto how information is stored in QCNN models, compared to their classical counterpart, and highlight the role of entanglement in this context. We plan to investigate this effect further and more rigorously in future work.   

\subsection{PTA: Post-training parameter adjustment}\label{subsec:paper}
Since setting (random, single-qubit) gates to the identity seemed to have a crucial effect on the network's performance, we hypothesized that tinkering with the trained parameters to a certain degree might provide enough flexibility to the model without hampering its accuracy and predicting capability. In the PTA method, rather than dropping out gates completely,  some of the trained parameters are slightly adjusted. The performance of the slightly modified model was then tested using testing and validation data.  

\begin{figure}[!ht]
  \centering
  \includegraphics[width=0.5\textwidth]{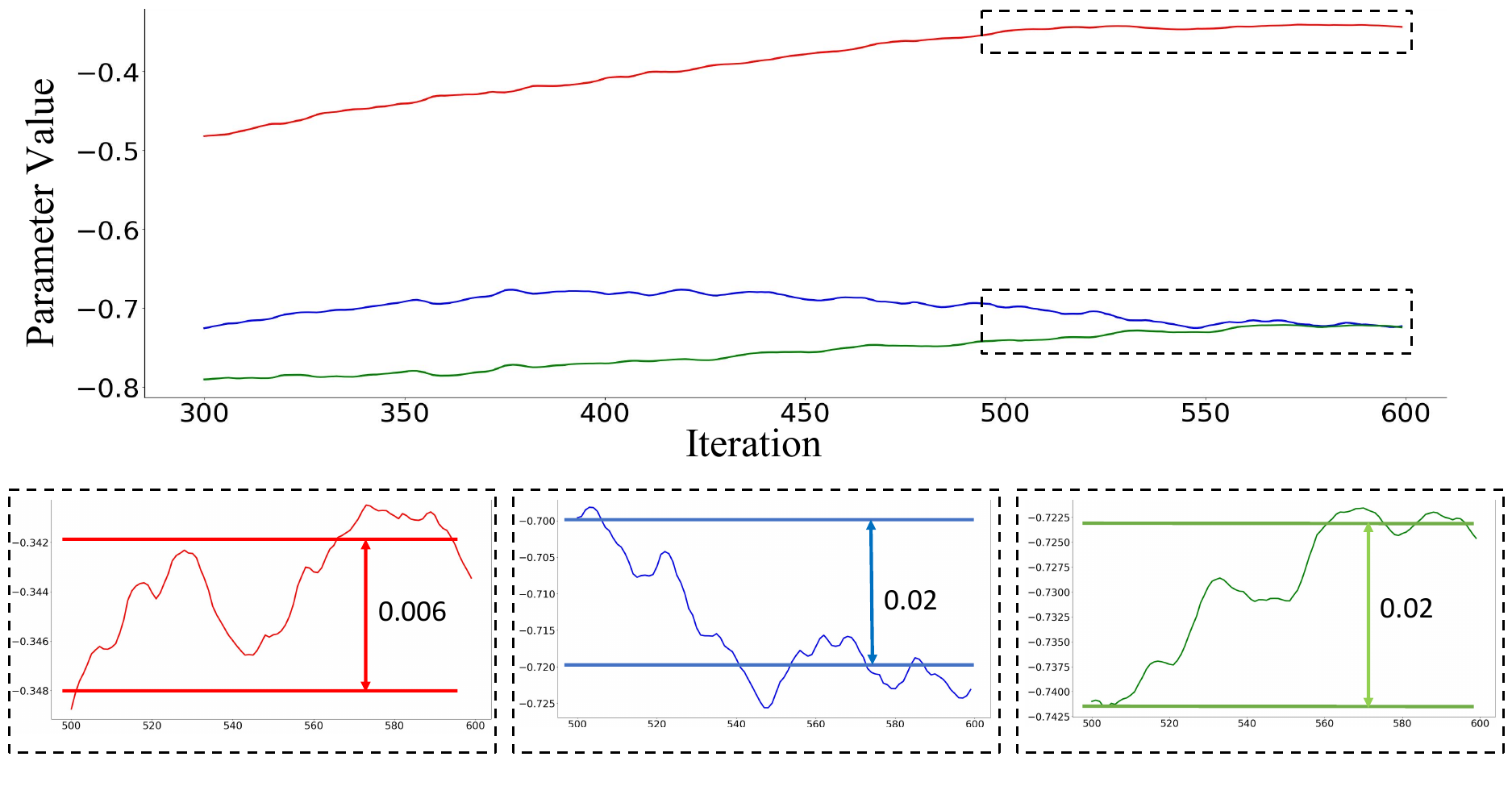}
  \caption{{\bf Parameters evolution over time (iterations).} The top plot shows how the values of three  (out of 78) parameters in one experiment we run (which were chosen for illustration purpose) change with the number of  iterations. The bottom row, shows a zoom on the last 100 iterations (color coded). In the PTA, we may choose to adjust the parameter value according to the range of its fluctuation in the last 100 iterations.}
  \label{fig:traces}
\end{figure}

There are several methods to choose the parameters to adjust and the how to adjust them using the PTA approach. For example, one can store the values of the parameters at the last few iterations of training the model (say, the last hundred iteration steps), and depending on their the fluctuation of their values, decide which parameters to  adjust and to what value (e.g., taking the mean of the last 100 iteration values for the chosen parameters).  In Fig.~\ref{fig:traces} we show an example of the values that three parameters take on while training a QCNN model on 8 qubits on the Medical MNIST data (see more detail below). The model, in this particular example, consists of a variational quantum circuit with 78 parameters, out of which three are shown for illustration purposes. As evident from the figure the values of the parameters have clear structure while training. This structure can be informative to decide which parameters to adjust and how. A compelling choice is to adjust the parameters to the mean value of the fluctuation towards the end of the training process.

To test the effectiveness of the PTA method, in this work we chose at random a small percentage ($1\%-5\%$) of the variational parameters (from the set of convolutional layers) and adjusted them manually. This allowed us to gauge, for the specific models we studied, the effect of the PTA method and the threshold at which the changing parameter falters. 
In particular we use adjustment techniques such as rounding the trained parameters up to certain decimal places, asserting certain values up to a threshold to a common whole number, and setting certain values close to zero, considering their proximity to the number. For the technique of setting values to zero, all the trained parameter values in the list of parameters below a certain floating number value (generally ranging below $\vert\pm0.09\vert$) were changed to float value $0$ by taking the absolute of all the parameters to cover all the values from the positive and negative spectrum of trained parameters. A similar technique was utilized in the case of threshold whole number conversion instead of setting it up to zero; the threshold depended on the point of mitigating overfitting without loss in accuracy dropping, which was found over several manual iterations of finding the least testing-validation accuracy value and not losing the actual testing value significantly. For the round-off method, a built-in Python function for rounding up values up to certain decimal places was used, while not all values were rounded up to certain decimal places as after a certain threshold, a severe drop in accuracy was observed. The threshold for the round-off method was determined by iterating the method until finding the  parameters that yield the highest validation accuracy compared to the unmitigated circuit. Results and observations of the PTA method are discussed for all the datasets in Sec.~\ref{sec:results}.

\section{Numerical experiments and results}\label{sec:results}
\subsection{Datasets}\label{subsec:data}
Using multiple datasets in ML, including QML, is very useful to increase generalization, robustness, and reliability. It also helps overcome data limitations, introduces variability and heterogeneity, and allows the exploration of different perspectives. 
In regards to these ideas, we chose to work with three datasets, two of them being image-based medical datasets, Medical MNIST~\cite{mnist} and BraTS~\cite{menze2015multimodal}, while the third was a Stellar dataset~\cite{sdss} consisting of numerical values. 

Each dataset was split into three parts: one for training, another for testing, and the last portion for validation. The validation set is used to test the performance of the (trained) QCNN on an unseen dataset and provide a proxy for determining if the model exhibits overfitting and how well we mitigate this problem.  Many variations of percentage split were implemented to find the best option for creating the overfitting conditions. This operation was performed for all three datasets. Later, after the split, the testing data was added to the training data to develop the overfitting conditions more prominently. 
Subsequently, the data were processed using PCA for dimensionality reduction and fitting it to the limited number of qubits (we used 8, 12, and 12 qubits). The classically-processed data was then sent for training on a (simulated) QCNN.

\begin{figure}[!ht]
  \centering
  \includegraphics[width=0.15\textwidth]{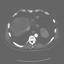}
  \includegraphics[width=0.15\textwidth]{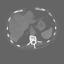}
  \includegraphics[width=0.15\textwidth]{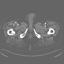}
  \includegraphics[width=0.15\textwidth]{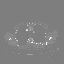}
  \includegraphics[width=0.15\textwidth]{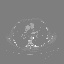}
  \includegraphics[width=0.15\textwidth]{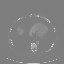}
  \caption{{\bf Example of Medical MNIST dataset images.} Top row: Abdominal CT, bottom row: Chest CT. The similarity between Chest CT and Abdominal CT images implies that they would be hard to classify and in addition, that the learning model may be prone to overfitting.}
  \label{fig:mnist}
\end{figure}

{\it Medical MNIST}.---~The Medical MNIST dataset consists of 6 classes: Abdominal CT, Breast MRI, Chest CT, Chest X-ray, Hand X-ray, and Head CT. These classes contained around 10000 images, each of size $64 \times 64$ pixels. As we are trying to implement binary classification, all possible permutations of classes were tested, and the similarity between Chest CT and Abdominal CT was considered most of the time. Several QCNNs were created to differentiate between the CT images being a Chest CT image or an Abdominal CT image. These two classes were pretty much alike and, hence, challenging to differentiate and were prone to overfitting, see Fig.~\ref{fig:mnist}.

\begin{figure}[!ht]
  \centering
  \includegraphics[width=0.15\textwidth]{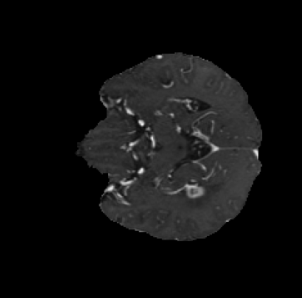}
  \includegraphics[width=0.15\textwidth]{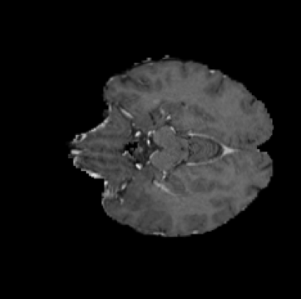}
  \includegraphics[width=0.15\textwidth]{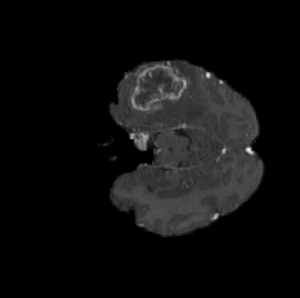}
  \includegraphics[width=0.15\textwidth]{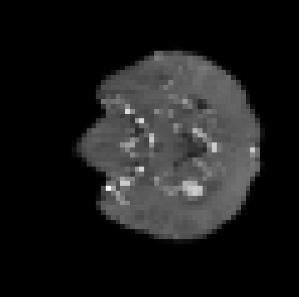}
  \includegraphics[width=0.15\textwidth]{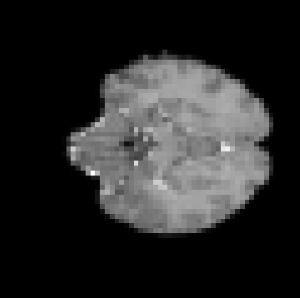}
  \includegraphics[width=0.15\textwidth]{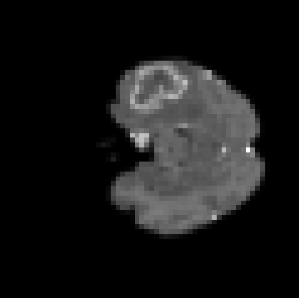}
  \caption{{\bf Example of BraTS 2019 dataset brain images} The high-resolution brain images at the top row were resized to 64 pixels, bottom row. The resulting images appeared unclear and pixelated, which pose a challenge for classification.}
  \label{fig:brats}
\end{figure}

{\it BraTS 2019}.---~To validate the uniformity of the proposed dropout approach on different datasets and models, we used another medical dataset for this implementation. The BraTS 2019 dataset  was chosen for classification between High-Grade Gliomas (HGG) and Low-Grade Gliomas (LGG) patient MRI images. The BraTS 2019 training dataset has 259 HGG and 76 LGG images. The BraTS multimodal scans are provided in NIfTI file format (.nii.gz) and include the following components: a) native (T1) scans, b) post-contrast T1-weighted (T1Gd) scans, c) T2-weighted (T2) scans, and d) T2 Fluid Attenuated Inversion Recovery (T2-FLAIR) scans. These scans were obtained using diverse clinical protocols and a variety of scanners from a total of 19 different institutions.
Due to resource limitations, only one modality, specifically the T2-FLAIR, was considered for the classification of HGG versus LGG. The images were resized to 64 pixels. As depicted in Fig.~\ref{fig:brats}, the resulting images appeared unclear and pixelated, which was expected given the constraints.

{\it Stellar classification dataset (SDSS17)}.---~As both of the prior datasets were image-based, we consider using a dataset in a different format to verify the conclusions devised and ascertain our claims derived from the results. The stellar classification dataset SDSS17 seemingly proved to be a reliable candidate. It consists of 100,000 observations of space taken from the Sloan Digital Sky Survey (SDSS).  Every recorded observation was registered in 17 columns defining different values, such as the ultraviolet filter in the photometric system, the green filter in the photometric system, the redshift value based on the increase in wavelength, etc., and one class column identifying the category for being a star, quasar or a galaxy. Out of the 17 available columns, only 8 were used for training the model after the initial data pre-processing (columns containing the ID of equipment and date parameters were removed in order to generalize the object regardless of its position in the sky and the equipment detecting it). Considering the close proximity of the star and quasar data and their difficulty in classifying based on the data available, we considered classifying these two classes using QCNN. As the data consisted of only 8 columns, PCA did not need to be applied to reduce the dimensionality of the data. Hence, all the experiments conducted for the classification of this dataset were limited to 8 qubits. The same process of data splitting was utilized as defined for the previous dataset with few exceptions for the train-test-validation split percentage in order to characterize the overfitting scenario.

\subsection{Results from numerical experiments}\label{subsec:result}
All the experimental data generated for this manuscript was generated using the Pennylane software (v$0.31$) and simulated on a local classical device, considering the total amount of iteration needed to train the QML model and the queuing operations required to complete the process on any of the available quantum computers on the cloud. For optimization purposes, we utilized the Pennylane optimizer Nesterov Momentum Optimizer, considering its merits observed during the initial training of the QML models~\cite{qmlNesterov}. The total number of qubits used for this experimentation varied from $8-16$ depending upon the dataset used and the complexity of the circuit. PCA, as previously mentioned in Sec.~\ref{sec:soft}, was utilized for converting higher dimensionality data onto the number of qubits defined for the QML model. 

We verify success in mitigating overfitting in our experiment using two figure of merits: (1) increasing validation accuracy and (2) reducing the difference between testing and validation accuracy after implementing the dropout method.

\subsubsection{Mitigating overfitting using post-training gate dropout}
The first method we implemented to mitigate overfitting in QCNN is a direct adaptation of (post-training) dropout to the quantum setting, as discussed above. We have applied this method to mitigate overfitting when considering the Medical MNIST, BraTS, and Stellar Classification datasets. We found that dropping out even a single gate (either a single-qubit gate or a CNOT gate) from the trained network had a devastating effect, regardless of the tested feature map and the dataset. For example, when we implemented this method on an 8-qubit  model that was trained on the Medical MNIST dataset with a success accuracy of $95\%\pm0.05\%$, by  dropping out  only $5\%$ of the single-qubit gates, the accuracy of the QCNN on the testing data was reduced significantly with $77\%$ accuracy being one of the best-performing models and about $2\%$ being the worst-performing over about a dozen of trained models.  Not only was this method not able to mitigate overfitting by increasing validation accuracy or reducing the gap between the testing and validation accuracy, but it also resulted in dramatically hampering the performance of the network on the trained data. Similar behavior was observed to be robust with respect to the number of single-qubit gates that were dropped out. In addition, we tested dropping out $1\%$ to $10\%$ of the single-qubit gates in a network that has 78 parameterized single-qubit gates, and observed a similar drastic drop in performance accuracy. This was contrary to our na{\"i}ve intuition that a network with many gates, more than the minimum required to accomplish the learning task,  should be minimally affected, if at all,  by dropping out a few single-qubit gates. To test the effect of the method at its limit, we implemented it by dropping out one (randomly chosen) single-qubit gate out of a model with 78 gates. This experiment resulted in very interesting results. In these cases, the accuracy was plunged to a range of about  $46\%$ to $53\%$ over a dozen of trained models, which is almost $50-50$ chance in the particular model we have tested. 

In addition to dropping out single-qubit gates, we tested post-training CNOT gate dropout. We found that by dropping (at random) a single CNOT gates from  trained models on Medical MNIST dataset, a tremendous loss of testing accuracy, from $94.3\%\pm0.05\%$ to $49.89\%\pm 0.05\%$, has been observed. This loss of accuracy persist (by $\pm 3\%$) while varying the building blocks for the convolutional layers. We saw similar results for the BraTS dataset, wherein the test accuracy decreased from $95.2\%$ to $65.5\%$  with fluctuations of about  $\pm0.07\%$  within about a dozen of  experiments. 

Lastly, in the case of the Stellar Classification dataset, the testing accuracy dropped from $94.0\%$ to $46.5\%$ ($\pm 0.07\%$) and validation accuracy dropped from $93.2\%$ to $30.5\%$ ($\pm 0.05\%$). 


This experimentation on post-training gate dropout bore the conclusion that even removing a single gate from the circuit of a trained QCNN may cause a dramatic loss of information gained during the training process, even for relatively deep PQCs. We hypothesize that entanglement plays a crucial role in this behavior. In classical CNN, where the post-training neuron dropout method is used very successfully, each neuron holds a few bits of information regarding the data it is trained on, and therefore, losing a small fraction of neurons does not affect the overall performance of the entire network. In stark contrast, QCNN is designed to harness entanglement between qubits in our implementation through the concatenation of single-qubit (parameterized) gates and CNOT gates. This means that the information learned about a certain dataset is stored and distributed in the QCNN in a ``non-local" way, loosely speaking. Our experiments indicate that entanglement may have a double-edged sword effect in QNNs: On one hand, it may promote speedup, e.g., in terms of learning rates, but on the other hand, it can lead to a fragile network, in the sense that removing even a single gate from a trained network may have devastating consequences with respect to its performance. This experiment, therefore, exposes an intrinsic vulnerability of QML, and QCNNs in particular, in comparison to their classical counterpart. We plan to study this effect more rigorously in a future work.

\begin{figure*}[!ht]
  \includegraphics[width=0.75\textwidth]{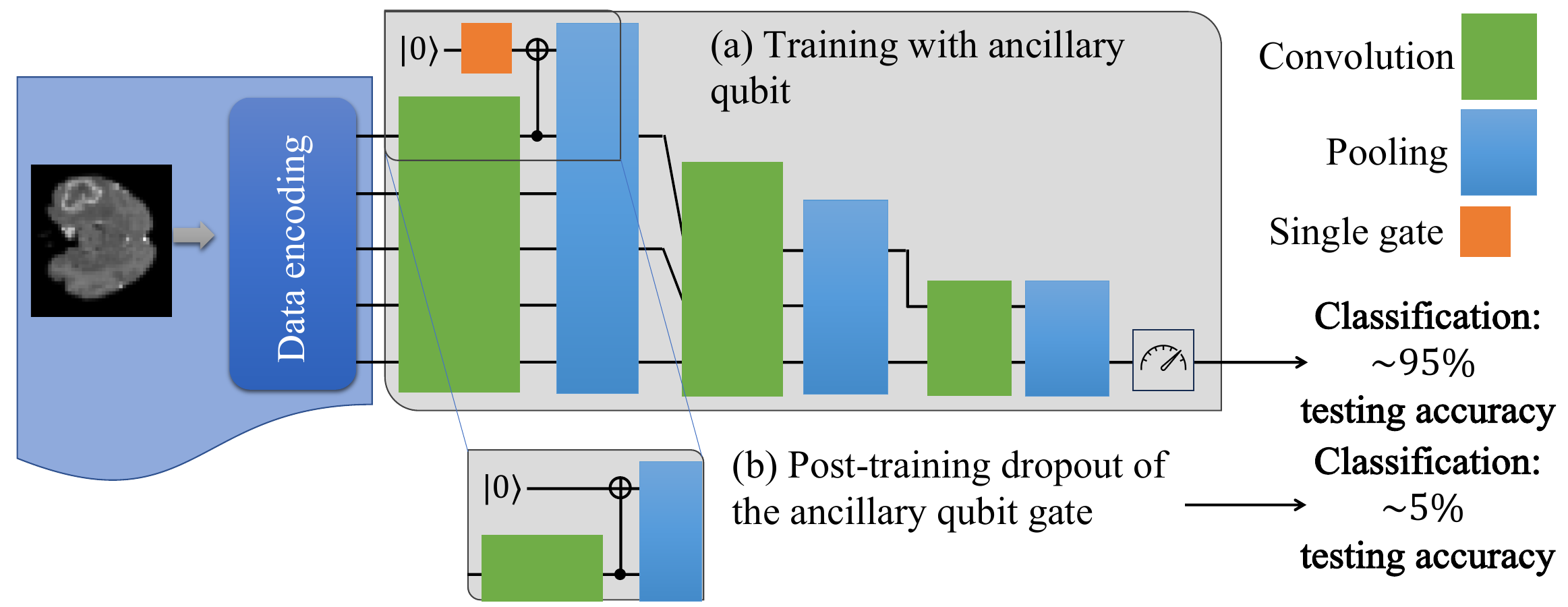}
  \caption{{\bf Ancillary qubit dropout experimental setup.} The figure  depicts the experimental setup used to exemplify the vulnerability of  QML models. (a) The QCNN is trained along with an ancillary qubit, which is not part of the feature map. The ancilla qubit takes part in the training via a parameterized rotation ($R_x$ or $R_y$). We found that this setup results in a testing accuracy of about 95\%. (b) After training is completed, removing the single-qubit gate from the ancillary qubit, the model experienced a significant loss in prediction accuracy.}
  \label{fig:dropout_exp}
\end{figure*}
To further ascertain the conclusion, we have conducted a set of experiments, schematically shown in Fig.~\ref{fig:dropout_exp}.  We have constructed a QCNN with 8 qubits and an additional ancillary qubit that does not pass through the feature map but rather is initialized in a computational state (say, $\ket{0}$) as a non-featured attestation to the QCNN circuit. Thus, this qubit does not hold any information about the input data. The ancillary qubit is then passed through a parameterized single-qubit gate (our experiments were done with an $R_x$ gate and a $R_y$ gate) whose parameters are consistently updated in every iteration of the training cycle along with the rest of the training parameters in the first convolutional layer. The qubit is then entangled with one of the qubits from the circuit with a CNOT gate after the first convolutional layer, and then it is traced out in the following pooling layer. Training this QCNN resulted in $93\%-95\%$ testing accuracy (depending on the network building blocks we used). However, by dropping out the parameterized gate of the ancillary qubit, the testing accuracy plunged to order of a few percents (between $2\%$ to $5\%$ in the tested models). This set of experiments clearly indicates that even though the ancillary qubit was not encoding information about the input data,  the mere fact that it is trained and entangled  with the rest of the qubits,  dropping after training, caused an information loss that resulted in a sharp accuracy drop. These results suggested that while dropping out gates in QCNN may not be a viable method for mitigating overfitting, tinkering with the trained values of the gates parameters may have a more subtle effect and thus can be used for this purpose.

\subsubsection{Mitigating overfitting using PTA}
As we discussed above, applying the classically derived method for post-trained gate dropout resulted in a drastic loss of learned information. In contrast, encouraging positive results were observed when the PTA method was applied. In these experiments we implemented the method by variations of rounding of the learned parameters and introducing a threshold on the values of the parameters, as prior mentioned in Sec.~\ref{sec:soft}.
 
We summarize our results in  Tables~\ref{tbl:medmnist result}-\ref{tbl:stellar result}, with respect to the datasets they are associated with. The results show the average test and validation accuracy over $8-10$ models, each of which the initial values of the parameters are set at random (in the range $[0,2\pi])$. The variation of the reported accuracy is $\pm 0.005$ for the Medical MNIST experiments, $\pm0.007$ for the BraST experiments, and $\pm0.04$ for the Stellar database experiments. The results clearly indicate that when a model suffers from overfitting (as captured by the lower validation accuracy and also an appreciable difference between test accuracy and validation accuracy), in most models that we have tested, the PTA method not only  was successful in reducing the gap between testing and validation accuracy but  also helped to increase the model validation accuracy across all of our experiments. 

\begin{table*}[htbp]
\begin{center}
\begin{tabular}{ c c c}
\multicolumn{3}{ c }{8 qubits} \\
\hline
             \textbf{Test Acc.} & \textbf{Validation Acc.} & {\bf Gap}  \\
            \hline 
             0.9154 & 0.7175 & 0.1979  \\
             \textcolor{blue}{0.9229} & \textcolor{blue}{0.8629} &\textcolor{blue}{0.06}  \\\hline
             0.9721 & 0.9136 & 0.0585 \\
            \textcolor{blue}{0.9794} &\textcolor{blue}{0.9447} &\textcolor{blue}{0.0347}\\ \hline
              0.9225 & 0.8770 & 0.0455\\
            \textcolor{blue}{0.9339}&\textcolor{blue}{0.9039} &\textcolor{blue}{0.03}\\ \hline
             0.9675 & 0.9298 & 0.0377\\
            \textcolor{red}{0.9464} &\textcolor{blue}{0.9374} &\textcolor{blue}{0.009}\\ 
            \hline
            \end{tabular}\quad\quad
\begin{tabular}{c c c}
\multicolumn{3}{ c }{12 qubits} \\
\hline
             \textbf{Test Acc.} & \textbf{Validation Acc.} & {\bf Gap}  \\
            \hline 
             0.9008 & 0.8866 & 0.0142  \\
            \textcolor{blue}{0.9545} &\textcolor{blue}{0.9079} &\textcolor{red}{0.0466}  \\\hline
             0.9560 & 0.9084 & 0.0476 \\
             \textcolor{blue}{0.9620} &\textcolor{blue}{0.9221} &\textcolor{blue}{0.0399} \\ \hline
              0.9345 & 0.9079 & 0.0266\\
             \textcolor{blue}{0.9360} &\textcolor{blue}{0.9085} &\textcolor{red}{0.0275}\\\hline
              0.8450 & 0.8380 & 0.0070\\
             \textcolor{blue}{0.8637} &\textcolor{blue}{0.8558} &\textcolor{red}{0.0079}\\ \hline
        \end{tabular}\vspace{5pt}
    \end{center}\vspace{-2em}
    \caption{{\bf Results based on Medical MNIST dataset.} The table shows the results for the test accuracy, validation accuracy, and the accuracy gap (the difference between the test accuracy and the validation accuracy) for models that were trained with qubits 8 and 12 (in the initial feature map layer). In each pair of rows, the top row (black font) indicates the average accuracy obtained from the trained model, while the bottom row (blue/red font) shows the corresponding accuracy when the PTA method was implemented. The average accuracy was obtained over about 10 models, each of which with a random initialization of the parameters, with the range of test and validation accuracy varies within $\pm0.005$ across the different models. Blue/red colour indicates success/failure in mitigating overfitting in the corresponding metric, respectively. Different pairs of rows correspond to training different models on the same datasets.  The results clearly indicate that the PTA method was able to improve validation accuracy on all tested cases and to shrink the accuracy gap by an order of magnitude for certain models. 
    }
    \label{tbl:medmnist result} 
\end{table*}

\begin{table*}[htbp]
\begin{center}
\begin{tabular}{ c c c}
\multicolumn{3}{ c }{8 qubits} \\
\hline
             \textbf{Test Acc.} & \textbf{Validation Acc.} & {\bf Gap}  \\
            \hline 
             0.8728 & 0.8548 & 0.018  \\
             \textcolor{blue}{0.8765} & \textcolor{blue}{0.8829} &\textcolor{blue}{-0.0064}  \\\hline
             0.8543 & 0.8499 & 0.0044  \\
             \textcolor{blue}{0.8655} & \textcolor{blue}{0.8757} &\textcolor{red}{0.0102}  \\\hline
             
            \end{tabular}\quad\quad
\begin{tabular}{c c c}
\multicolumn{3}{ c }{12 qubits} \\
\hline
             \textbf{Test Acc.} & \textbf{Validation Acc.} & {\bf Gap}  \\
            \hline 
             0.8958 & 0.8859 & 0.01  \\
            \textcolor{blue}{0.8996} &\textcolor{blue}{0.9082} &\textcolor{blue}{-0.0086}  \\\hline
            0.8666 & 0.8645 & 0.0021  \\
            \textcolor{blue}{0.8731} &\textcolor{blue}{0.8793} &\textcolor{red}{0.0062}  \\\hline
        \end{tabular}\quad\quad
\begin{tabular}{c c c}
\multicolumn{3}{ c }{16 qubits} \\
\hline
             \textbf{Test Acc.} & \textbf{Validation Acc.} & {\bf Gap}  \\
            \hline 
             0.9422 & 0.9257 & 0.0165  \\
            \textcolor{blue}{0.9518} &\textcolor{blue}{0.9586} &\textcolor{blue}{-0.0068}  \\\hline
             0.8972 & 0.8895 & 0.0077  \\
            \textcolor{blue}{0.9127} &\textcolor{blue}{0.9233} &\textcolor{red}{0.0106}  \\\hline
        \end{tabular}\vspace{5pt}
    \end{center}\vspace{-2em}
    \caption{{\bf Results based on BraTS dataset.} The format of results is similar to Table~\ref{tbl:medmnist result}, for different models that were trained with qubits 8, 12 and 16. The results indicate an average over $8-10$ experiments, that correspond to a random initialization of the parameters. The accuracy range fluctuates between $\pm0.005$ to $\pm0.007$ for both test and validation sets.    
    For all experiments, the validation accuracy after  PTA was implemented is higher on average than without it. This suggests that PTA regularization technique helps improve the model's generalization performance and reduces overfitting.}
    \label{tbl:brats result}
\end{table*}

\begin{table}[htbp]
\begin{center}
\begin{tabular}{ c c c}
\multicolumn{3}{ c }{8 qubits} \\
\hline
             \textbf{Test Acc.} & \textbf{Validation Acc.} & {\bf Gap}  \\
            \hline 
             0.8866 & 0.8432 & 0.0434  \\
             \textcolor{blue}{0.8890} & \textcolor{blue}{0.8542} &\textcolor{blue}{0.0348}  \\\hline
             0.9456 & 0.9251 & 0.0165 \\
            {0.9456} &\textcolor{blue}{0.9399} &\textcolor{blue}{0.0057}\\ \hline
             0.9500 & 0.9225 & 0.0275\\
            \textcolor{red}{0.9455}&\textcolor{blue}{0.9400} &\textcolor{blue}{0.0055}\\
\hline
\end{tabular}
\end{center}
 \caption{{\bf Results based on Stellar dataset.}  Results have the same format as in Table~\ref{tbl:medmnist result}. Variations in accuracy for testing and validation fluctuate by $\pm0.003$ and $\pm0.004$, respectively, but the difference (gap) stays approximately the same for the same unitary structure. The results clearly indicate that PTA was successful in mitigating overfitting, as indicated by higher validation accuracy and smaller gap, as compared to the results where PTA was not implemented, across all tested models.}
    \label{tbl:stellar result}
\end{table}
We attempted to devise a systematic way to determine the threshold for rounding up to a number or an absolute value that could be decided for mitigating overfitting after obtaining the trained parameters. Utilizing this method on several trained and overfitted models, we observed that every model had a different threshold which could only be determined after constant testing to find the best fit for tackling the overfitting issue. In addition, a closer observation revealed that the set of parameters which were used to successfully mitigate  overfitting were those which fluctuated around a mean value and did not changed towards the end of the training process.  This observation will be explored in more detail in future work.

\section{Conclusion and Outlook}\label{sec: Conclusion}
In this study we focus on addressing the challenge of overfitting in QML setting, specifically in QCNNs.  Overfitting, a common issue in ML models, poses significant obstacles to the generalization performance of QCNNs. To overcome this challenge, we introduced and explored the potential of the PTA method and compared it to a straightforward application of a post-training dropout method  commonly utilized in classical CNNs~\cite{gal16dropout}.  Surprisingly, we found that dropping out even a single parameterized gate or a single CNOT gate from a trained QCNN can result in a dramatic decrease in its performance. This result highlights a vulnerability of QCNNs compared to their classical counterparts. On the other hand the PTA approach resulted in encouraging results on the tested models.

Extensive experimentation is conducted on diverse datasets, including Medical MNIST, BraTS, and Stellar Classification, to evaluate the effectiveness of PTA approach in mitigating overfitting in QCNNs.  Our findings highlight the promising performance of PTA in reducing overfitting and enhancing the generalization capabilities of QCNN models, irrespective of the  feature map encoding scheme, and convolutional circuit building blocks which were used in the experiments. By adjusting the trained parameters through various techniques, notable improvements in accuracy are observed while preserving the integrity of the quantum circuit. Hence, based on these findings we promote PTA as a viable option to mitigate overfitting in a post-training setting, in conjunction to pre- and during-training techniques.

We close this section with a few directions for future work. The first direction is developing a systematic approach for determining which, and how, parameters should be tinkered to handle overfitting. Following our initial observation, we believe that identifying those parameters that fluctuate around a mean value during training play an important role for mitigating overfitting.  

Another important direction for future work is to investigate the performance of PTA in the presence of experimental noise. Quantum systems are inherently susceptible to noise, which can impact the reliability and effectiveness of quantum operations. Understanding how PTA performs under noisy conditions will contribute to the development of robust QCNN models that can operate in realistic quantum computing environments.

Another aspect that requires further exploration is the scalability and performance of PTA in larger QCNN models. As quantum hardware continues to advance, larger and more complex QCNN architectures become feasible. Evaluating the behavior and effectiveness of PTA in handling larger quantum circuits will provide insights into its scalability and potential challenges in maintaining regularization benefits.

By pursuing these research directions, we hope to advance the field of QML and enhance the practical deployment of QCNN models. Overcoming overfitting challenges is crucial for ensuring the reliability and effectiveness of QCNNs in real-world applications, unlocking their potential to make significant contributions in various domains.

\begin{acknowledgments}
This project was supported in part by NSF award \#2210374. The code for all of the experiments presented in this work can be provided upon a reasonable request from the authors.

\end{acknowledgments}

\bibliography{biblio}
\end{document}